\documentclass{pasj00}
\draft

\begin{document}
\SetRunningHead{Y. Takeda et al.}{Volatile and Refractory Elements 
Abundances in Planet-Harboring Stars}
\Received{}%{yyyy/mm/dd}
\Accepted{}%{yyyy/mm/dd}

\title{
Photospheric Abundances of Volatile and Refractory Elements in 
Planet-Harboring Stars
}

%%% begin:list of authors
\author{
   Yoichi \textsc{Takeda},\altaffilmark{1}
   Bun'ei \textsc{Sato},\altaffilmark{2,3}
   Eiji \textsc{Kambe},\altaffilmark{4}
   Wako \textsc{Aoki},\altaffilmark{5}\\
   Satoshi \textsc{Honda},\altaffilmark{5}
   Satoshi \textsc{Kawanomoto},\altaffilmark{5}
   Seiji \textsc{Masuda}, \altaffilmark{6}
   Hideyuki \textsc{Izumiura},\altaffilmark{2}\\
   Etsuji \textsc{Watanabe},\altaffilmark{2}
   Hisashi \textsc{Koyano},\altaffilmark{2}
   Hideo \textsc{Maehara},\altaffilmark{2}
   Yuji \textsc{Norimoto},\altaffilmark{2}\\
   Takafumi \textsc{Okada},\altaffilmark{2}
   Yasuhiro \textsc{Shimizu},\altaffilmark{2}
   Fumihiro \textsc{Uraguchi},\altaffilmark{2}
   Kenshi \textsc{Yanagisawa},\altaffilmark{2}\\
   Michitoshi \textsc{Yoshida},\altaffilmark{2}
   Shoken \textsc{Miyama},\altaffilmark{5}
   and
   Hiroyasu \textsc{Ando}\altaffilmark{7}
   }
 \altaffiltext{1}{Komazawa University, Komazawa, Setagaya, Tokyo 154-8525}
 \email{takedayi@cc.nao.ac.jp}
 \altaffiltext{2}{Okayama Astrophysical Observatory, National Astronomical 
 Observatory,\\ Kamogata, Okayama 719-0232}
 \altaffiltext{3}{Department of Astronomy, School of Science,\\
 The University of Tokyo, Bunkyo-ku,Tokyo 113-0033}
 \altaffiltext{4}{National Defence Academy, Yokosuka, Kanagawa 239-8686}
 \altaffiltext{5}{National Astronomical Observatory, Mitaka, Tokyo 181-8588}
 \altaffiltext{6}{Department of Astronomy, Faculty of Science, 
 Kyoto University, Sakyo-ku, Kyoto 606-8502} 
 \altaffiltext{7}{Subaru Telescope, National Astronomical 
 Observatory of Japan,\\ 650 North A'ohoku Place, Hilo, HI 96720, U.S.A.}
%%% end:list of authors

%%% Please use the following style in case that sorting by 
%%% affilation is impossible. 
%
% \author{%
%   D-Firstname \textsc{D-Familyname}\altaffilmark{1}
%   E-Firstname \textsc{E-Familyname}\altaffilmark{1,2}
%   and
%   F-Firstname \textsc{F-Familyname}\altaffilmark{2}}
% \altaffiltext{1}{Address of Institute}
% \email{ddddd@xxx.xxx.xx.xx}
% \email{eeeee@xxx.xxx.xx.xx}
% \altaffiltext{2}{Address of Institute}

%% `\KeyWords{}' always has to be placed before `\maketitle'.
%\KeyWords{xxxx:xxxx ......} %Do NOT move this preamble from here!

\maketitle

\begin{abstract}
By using the high-dispersion spectra of 14 bright planet-harboring stars
(along with 4 reference stars) observed with the new coud\'{e} echelle 
spectrograph at Okayama Astrophysical Observatory, we investigated
the abundances of volatile elements (C, N, O, S, Zn; low condensation 
temperature $T_{\rm c}$) in order to examine whether these show any 
significant difference compared to the abundances of other refractory 
elements (Si, Ti, V, Fe, Co, Ni; high $T_{\rm c}$) which are known to be
generally overabundant in those stars with planets, since a
$T_{\rm c}$-dependence is expected if the cause of such a metal-richness 
is due to the accretion of solid planetesimals onto the host star.
We found, however, that all elements we studied behave themselves 
quite similarly to Fe (i.e., [X/Fe]$\simeq 0$) even for the case of 
volatile elements, which may suggest that the enhanced 
metallicity in those planet-bearing stars is not so much an acquired
character (by accretion of rocky material) as rather primordial.
\end{abstract}

%\section{}
%
%\noindent IMPORTANT NOTICE\\
%1. ``\verb|\draft|'' creates single column and double spaces format.\\
%2. If you comment out ``\verb|\draft|'', the output will be double column
%   and single space.\\
%3. For cross-references, the use of ``\verb|\label|, \verb|\ref|, \verb|\cite|'%' 
%   and the thebibliography environment is strongly recommended. \\
%4. Do NOT use ``\verb|\def|, \verb|\renewcommand|''.\\
%5. Do NOT redifine commands provided by PASJ00.cls.\\

%\newpage

\section{Introduction}

The topic of this paper concerns the chemistry of
planet-harboring solar-type stars, which have recently been
(and are now being) discovered one after another mostly by the Doppler 
technique thanks to the energetic activities of several groups. 

It seems to have been almost established observationally 
that those stars bearing giant planets tend to be mildly metal-rich 
(i.e., $\sim$+0.2 dex on the average with a considerable scatter) 
compared to the Sun or similar standard stars, though there are
some exceptions (see, e.g., Gonzalez et al. 2001; Sadakane et al. 2001,
Santos et al. 2001; Smith et al. 2001; and references therein).  
What is now in hot debate is, however, the {\it cause} of 
this tendency --- why does the existence of planets have something
to do with the photospheric abundances of host stars?

Here, a promising and often invoked interpretation is
the ``self-enrichment'' hypothesis. This attributes the origin of
the overabundance of metals to the accretion of large amount of 
H-depleted (i.e., relatively metal-rich) rocky planetesimal materials 
onto the star, which is more or less reasonably understood from 
the viewpoint of the planet-formation process.

Following this line, theoretical as well as observational investigations 
have recently been done to clarify whether this mechanism accounts for 
the observed abundance characteristics also in the quantitative sense.
Murray and Chaboyer (2001) arrived at a kind of affirmative conclusion
that adding $\sim$6.5 $M_{\oplus}$ of iron to each star can explain
both the mass--metallicity and the age--metallicity relations
of the stars with planets.
Contrary to this, however, Pinsonneault et al. (2001) pointed out that 
an evident $T_{\rm eff}$-dependence (i.e., more enhanced metallicity in 
earlier-type stars) must be observed if such an accretion mechanism 
is really responsible, which is not observed.
Similarly, Santos et al. (2001) concluded by examining the shape
of the metallicity distribution that the ``pollution scenario''
due to accretion is unlikely. They attributed the higher metallicity of 
planet-harboring stars to the simple fact that higher gas metallicity 
leads to the high probability of planet formation.

Meanwhile, another important key to check the ``self-enrichment'' 
hypothesis is to study the abundance pattern of various elements, 
especially the difference between the volatile elements (such as C, N, O) 
with low condensation temperature $T_{\rm c}$ (i.e., rather unlikely to 
be comprised in rocky materials) and the refractory elements (such as 
the iron-group elements) with high $T_{\rm c}$ (being apt to be condensed 
into solid particles). Namely, since the elements of the former group 
are expected to be deficient in the accreted materials relative to 
the latter, the abundances of the volatile group would not show 
such an overabundance as exhibited by the refractory group (e.g., Fe).

Then, is such a trend really observed? 
Sadakane et al.'s (1999) result for HD 217107, that C and O 
do not appear to share the tendency of marked overabundance shown by 
the other elements, may be suggestive for this prediction.
Also, Gonzales and Laws (2000) once stated that planet-harboring 
stars with [Fe/H]$>0$ tend to be [C/Fe]$<0$, again suggesting that
C does not keep pace with the enrichment of Fe (though other
stars without planets appear to show the similar tendency;
see also Santos et al. 2000).
Later, however, Gonzalez et al. (2001) withdrawed this claim 
by stating ``that stars with planets have low [C/Fe] is found to be
spurious, due to unrecognized systematic differences among published
studies''.
On the other hand, Smith et al. (2001) found that a small subset of
planet-harboring stars (i.e., those with large planets orbiting 
quite close) exhibit a trend of increasing [X/H] with
an increase in $T_{\rm c}$, which is a sign of the accretion of 
chemically fractionated solid material.

Thus, the situation is rather confusing, and a new systematic
observational study is required to settle this problem.
Following this motivation, we decided to investigate,
based on our own observational data of bright planet-bearing stars
collected at Okayama Astrophysical Observatory, whether any difference 
or systematic tendency exists between the abundances of volatile and 
refractory elements, which can be an important touchstone for
the adequacy of the ``self-pollution'' scenario.
This is the purpose of the present study, in which
we will focus on the abundances of five volatile 
elements (C, N, O, S, and Zn) and compare them with those of
the refractory elements (Si, Ti, V, Fe, Co, and Ni)

\section{Observational Data}

We observed 18 stars (14 planet-harboring stars and 4 reference stars)
listed in table 1 by using the new high dispersion echelle specstrograph 
(HIDES; cf. Izumiura 1999) at the coud\'{e} focus of the 188~cm reflector 
of Okayama Astrophysical Observatory in 2000 April, 2000 October, 
2000 December, 2001 February, and 2001 March.

The slit width was set to 200~$\mu$m ($0.''76$), by which 
the spectral resolution of $R \sim 70000$ was accomplished.
By using the single 4K$\times$2K CCD (13.5~$\mu$m pixel),
the wavelength span of $\sim 1200\rm\AA$ could be covered at one time.
We have chosen three wavelength regions: green-centered region 
(4900--6100 $\rm\AA$; G), red-centered region (5900--7100 $\rm\AA$; R),
and (near-)IR-centered region (7600--8800 $\rm\AA$; I).
All stars could be observed at both of the wavelength regions G and R, 
while the region I data could not be obtained for six stars 
($\epsilon$ Eri, 51 Peg, HD 217107, 14 Her, $\tau$ Cet, and $\beta$ Vir).

The reduction of echelle data (bias subtraction, flat-fielding, 
scattered-light subtraction, spectrum extraction, 
and wavelength calibration) was performed by using 
the IRAF\footnote{IRAF is distributed
    by the National Optical Astronomy Observatories,
    which is operated by the Association of Universities for Research
    in Astronomy, Inc. under cooperative agreement with
    the National Science Foundation.} 
software package in a standard manner.

The S/N ratios of the resulting spectra differ from 
case to case, but are typically of the order of 200--300.
However, the near-IR spectra turned out to be generally 
of poorer quality because interference fringes could not 
completely be removed.

Regarding the solar spectrum used as a reference standard,
we adopted the very high resolution and very high S/N 
solar flux spectrum published by Kurucz et al. (1984).

\section{Abundance Analysis}

\subsection{Model Atmospheres}

With regard to the atmospheric parameters ($T_{\rm eff}$, $\log g$,
[Fe/H], and $\xi$) of the program stars necessary for 
constructing model atmospheres and determining abundances,
various published studies were consulted.
We compared the published values from several sources 
(e.g., Gonzalez 1998; Gonzalez et al. 2001; Gonzalez, Law 2001; 
Fuhrmann 1998) for a same star when available, and found that 
they are generally consistent with each other and that 
the differences are in most cases within 
$\Delta T_{\rm eff}\sim $100~K and $\Delta\log g \sim$0.1. 
The adopted parameter values (with the references) are presented 
in table 1.

We interpolated Kurucz's (1993) grid of ATLAS9 model atmospheres
in terms of $T_{\rm eff}$, $\log g$, and [Fe/H] to generate
the atmospheric model for each star.

\subsection{Determination of the Abundances}

\subsubsection{Equivalent width analysis}

Regarding the target volatile elements (C, N, O, S, and Zn), 
as far as the lines of interest are neither blended nor too weak, 
we determined the abundances from the equivalent widths by using 
the WIDTH9 program, a companion program to the ATLAS9 model 
atmosphere program written by Dr. R. L. Kurucz (Kurucz 1993).
As for the data of spectral lines (wavelengths, $gf$ values, and 
damping parameters), we exclusively invoked the compilation
of Kurucz and Bell (1995). 
The $gf$ values of important lines relevant to our analyses
are presented in table 2, though the choice of absolute oscillator
strengths is not essential because the analysis of this study is 
differential relative to the Sun.
The observed equivalent widths (which were measured by the method of 
Gaussian fitting or direct integration depending on cases) and the
resulting abundances are given in tables 3 and 4, respectively.

Note that the S abundance from the S~{\sc i} 6052 line was
determined by correctly taking into account the blending two components,
while the equivalent width was measured as if regarding it as a
single line.
Also, the non-LTE correction (amounting to $\sim$0.1--0.4 dex) 
was taken into consideration for deriving the O abundance 
from the strong O~{\sc i} 7771, 7774, and 7775 triplet lines according 
to the calculation of Takeda (2001), as given in the 7th column of
table 4.

\subsubsection{Spectral synthesis analysis}

Meanwhile, refractory element abundances (Si, Ti, V, Fe, Co, and Ni)
were determined by the spectrum synthetic technique while applying
the automatic fitting algorithm (Takeda 1995), 
in which the abundances of these 6 elements (along with 
the macrobroadening parameter and the radial velocity shift) 
were simultaneously varied to find the best-fit solution 
such that minimizing the $O-C$ deviation.
This yields accurate and reliable results than the case of conventional 
analysis. After several trials, the 6080--6089 $\rm\AA$
region was found to be adequate (i.e., a satisfactory fit is
accomplished with our $gf$ values given in table 2), which we decided 
to adopt. Note that, apart from those given in table 2, all lines 
listed in Kurucz and Bell's (1995) compilation were included
as background opacities.
The best-fit theoretically synthesized spectra along with the observed
spectra are shown in figures 1a and b.
Such established abundances are given in table 4.

Similarly, the C and O abundances from [C~{\sc i}] 8727 and O~{\sc i} 6158 
lines were obtained by this technique, since these line features are
delicately weak and more or less blended with other species.
As a result of the analyses of these two line features, the abundances 
of Si and Fe were simultaneously obtained as by-products.
Note that spectral portions, where the line data are so insufficient that
the observed spectrum could not be reproduced
(e.g.,around 6157.3 $\rm\AA$ for the case of lower temperature stars), 
were masked (i.e., excluded from judging the goodness of fit) 
in the process of automatic fitting.
Figures 2 (8727~$\rm\AA$ region analysys) and 3 (6157~$\rm\AA$ region 
analysis) show the appearance of the fit, and the resulting abundances are
presented in table 4.

\subsubsection{Sensitivity to parameter changes}

In order to estimate the uncertainties involved with the
ambiguities in the atmospheric parameters, we selected $\upsilon$ And
and 70 Vir as the respresentatives of higher $T_{\rm eff}$ and lower
$T_{\rm eff}$ stars and evaluated the abundance variations in response
to changing $T_{\rm eff}$ (by +150~K), $\log g$ (by +0.3), and $\xi$
(by +0.5~km~s$^{-1}$).
The results are given in table 5, from which we may state that errors
caused by uncertainties in these parameters are not very significant
considering the reasonable consistency of the published values 
(cf. subsection 3.1).

\subsubsection{Final results}

In table 6, we give the final results of the abundances
which will be discussed in section 4.

As for the volatile elements, we adopted a mean value when two or 
more abundances from different line indicators are available. 
In this averaging process, a half weight was assigned to the C and O
abundance derived from the forbidden lines ([C~{\sc i}] 8727
and [O~{\sc i}] 6300) in view of their comparatively less reliability
due to the suspected blending effect, i.e., with unknown species 
for [C~{\sc i}] (Gonzalez et al. 2001) and with the Ni~{\sc i} line 
for [O~{\sc i}] (Allende Prieto et al. 2001).

Meanwhile, regarding the refractory elements (Si, Ti, V, Fe, Co, and Ni),
we adopted the abundances from the 6080--6089 $\rm\AA$
fitting analysis, since we consider the results for these 6 elements
from this 6080--6089 $\rm\AA$ region is sufficiently accurate.

The $T_{\rm c}$ values presented in table 6 were taken from
Field (1974) for CNO, and from Wasson (1985) for the remaining 
elements.

\section{Discussion}

\subsection{Fe abundance}

The resulting Fe abundances of the program stars (table 6) are 
compared with the literaure values (table 1) in figure 4,
where we can see that both are in fairly reasonable agreement.
The mean for those 14 stars with planets is 
$\langle$[Fe/H]$\rangle = +0.12 (\pm 0.20)$,
which confirms the mild enrichment of metallicity as has been
already reported (cf. section 1), though a rather large diversity
is observed (e.g., $\rho$ CrB is evidently metal-deficient,
though its companion may actually not be a planet as remarked
in the footnote to table 1).

\subsection{Refractory Elements}

Figures 5a--e show the comparison of the abundances of five
refractory elements (Si, Ti, V, Co, and Ni) with the Fe abundance.
We may state from this figure that the abundances of those elements 
scale with that of Fe, which is an expected tendency because these
belong to the same group in terms of $T_{\rm c}$ (1300--1500 K).

The reason for the appreciable systematic difference of [Co/H] by 
0.1--0.2 dex relative to [Fe/H] (cf. figure 5d) may be related to 
the weakness of this line (especially for higher $T_{\rm eff}$ stars; 
cf. figure 1a), by which the reliability of the results may be 
more or less lowered.

\subsection{Volatile Elements}

The results of [C/H], [N/H], [O/H], [S/H], and [Zn/H] are
compared with [Fe/H] in figures 6a-e, respectively.
These figures suggest that {\it the abundances of all these five 
volatile elements scale with that of Fe}.
\footnote{The trend of appreciable deviation for $\tau$ Cet (observed 
also in figure 5) is (at least partly) attributed to 
the consequence of the Galactic chemical evolution for $\alpha$-process 
elements (e.g., O, S, Si, and Ti), for which metal-deficient stars 
([Fe/H]~$<$~0) are known to show [$\alpha$/Fe]~$>$~0. However, the reason 
why this tendency more or less appears in other (non-$\alpha$) 
elements is not clear.}
Especially, the trend of [C/Fe]~$<$~0 for metal-rich
planet-harboring stars, which was once reported before (cf. section 1),
can not be confirmed. 
Hence, our results suggest that the tendency predicted from 
the ``self-enrichment'' hypothesis, [X/H]~$<$~[Fe/H] for metal-enriched 
planet-harboring stars ([Fe/H]~$>$~0), is not observed at all.
(Even a weak inversed trend, [X/H]~$>$~[Fe/H], appears to exist
such as for N or S.)

If fractionated solid materials have something to do with any 
chemical peculiarity in photospheric abundances, some kind of 
anti-correlation between the abundances of voilatile and refractory 
elements is to be expected. Let us consider, for example, the case
of A-type stars (including $\lambda$ Boo stars). In this class
of stars, [C/H] is known to be clearly anti-correlated with [Si/H],
which is suspected as due to the differential accretion of gas and dust
in late stages of star formation (see, e.g., Holweger, St\"urenburg 1993).
On the contrary, in the present case, C and Si behave themselves 
quite similarly with each other, as shown in figure 7 
(compare this figure with their figure 2).

We also investigated if there is any systematic trend in [X/H]
with respect to $T_{\rm c}$, such that reported by Smith et al. (2001) 
for a subset of planet-harboring stars.
We calculated the slope ($a$) of the linear-regression line 
([X/H]=$a \cdot T_{\rm c} + b$) determined from the [X/H] values 
of the 11 elements (or 10 elements when N is not available).
The results are given in 13th and 14th columns of table 6.
By inspecting these $a$ values, the signs of which are randomly
positive or negative without showing any systematic dependence 
with the orbital period/radius of the planets, we may state that 
the tendency of positive $a$ expected from the self-enrichment 
hypothesis (cf. section 1) can not be observed.

Consequently, we consider that the ``pollution'' scenario is 
rather unlikely for the explanation of the metal-richness of 
planet-harboring stars, which may lend support for an alternative
interpretation that the enhanced metallicity of stars with planets 
is nothing but a ``primordial'' characteristics.

\section{Conclusion}

We determined the abundances of five volatile elements (C, N, O, S, and Zn;
low $T_{\rm c}$) and six refractory elements (Si, Ti, V, Fe, Co, and Ni;
high $T_{\rm c}$) for 14 bright planet-bearing stars along with
4 reference stars (+ Sun), based on the spectra obtained by using
the high-dispesion echelle spectrograph at Okayama Astrophysical
Observatory, in order to examine whether any systematic difference
exists between these two groups of elements.

Our results indicate that all of the studied elements behave themselves
quite similarly with each other (i.e., [X/Fe]$\simeq 0$) irrespective
of whether the element is volatile (apt to be remained in gas phase) 
or refractory (apt to be included in solid materials).
In other words, we could not observe any systematic 
tendency of [X/H] in terms of $T_{\rm c}$.

These facts suggest that the ``self-pollution'' scenario
proposed to explain the enhanced metallicity in planet-harboring stars,
which invokes an accretion of metal-rich solid planetesimals
(deficient in volatile elements while rich in refractory ones) onto
the host star, is rather unlikely, since this mechanism would
result in [X/Fe]~$<$~0 for [Fe/H]~$>$~0 for volatile elements
and a systematic increase of [X/H] with $T_{\rm c}$.

Therefore, as far as those 14 bright sample stars studied in this
investigation are concerned, we would consider that the tendency of 
metal-richness in planet-bearing stars is of primordial origin 
(i.e., higher probability of planet formation for intrinsically 
metal-rich gas) rather than attributing it to an acquired character.
\newline
\newline
This research have made use of the simbad database, operated at 
CDS, Strasburg, France.
This work is based on observations carried out within the framework of 
the OAO key project ``Comprehensive Spectroscopic Study of 
Stars with Planets'', which aims to understand the property of 
planet-harboring stars by analyzing stellar abundances and line-profles, 
as well as to improve the precision of measuring stellar radial 
velocities to a level of planet-detectability.

\clearpage
\setcounter{table}{0}
\begin{table}[h]
\footnotesize
\caption{Data and adopted atmospheric parameters of the observed stars.}
\begin{center}
\begin{tabular}{
r@{ }c@{ }c@{ }c
c@{ }c@{ }c
r@{ }r@{ }r@{ }c@{ }c}\hline\hline
HD No. & Name & $V$ & Sp & $M\sin i$ & $a$ & $P$ & $T_{\rm eff}$ & $\log g$ & [Fe/H] & $\xi$ & Reference \\
 &  &  &  & ($M_{\rm J}$) & (AU) & (d) & (K) & (cgs) &  & (km~s$^{-1}$) &  \\ \hline
120136 & $\tau$ Boo & 4.50 & F6IV & 3.87 & 0.05 & 3.31 & 6420 & 4.18  & +0.32  & 1.3  & GL \\
9826 & $\upsilon$ And & 4.09 & F8V & 0.71 & 0.06 & 4.62 & 6140 & 4.12  & +0.12  & 1.4  & GL \\
186427 & 16 Cyg B & 6.20 & G2.5V & 1.5 & 1.72 & 804 & 5700 & 4.35  & +0.06  & 1.0  & G98 \\
22049 & $\epsilon$ Eri$^{*}$ & 3.73 & K2V & 0.86 & 3.3 & 2502.1 & 5180 & 4.75  & $-$0.09  & 1.3  & DS \\
95128 & 47 UMa & 5.10 & G1V & 2.4 & 2.11 & 1082.02 & 5800 & 4.25  & +0.01  & 1.0  & G98 \\
89744 & HD 89744 & 5.74 & F7V & 7.2 & 0.88 & 256 & 6338 & 4.17  & +0.30  & 1.6  & GLTR \\
217014 & 51 Peg & 5.49 & G2.5IVa & 0.45 & 0.05 & 4.23 & 5750 & 4.40  & +0.21  & 1.0  & G98 \\
75732 & 55 Cnc & 5.95 & G8V & 0.84 & 0.11 & 14.65 & 5250 & 4.40  & +0.45  & 0.8  & GV \\
117176 & 70 Vir & 5.00 & G4V & 6.6 & 0.43 & 116 & 5500 & 3.90  & $-$0.03  & 1.0  & G98 \\
143761 & $\rho$ CrB$^{\dagger}$ & 5.40 & G0Va & 1.1 & 0.23 & 39.65 & 5750 & 4.10  & $-$0.29  & 1.2  & G98 \\
217107 & HD 217107 & 6.18 & G8IV & 1.28 & 0.07 & 7.13 & 5600 & 4.40  & +0.36  & 1.0  & GLTR \\
52265 & HD 52265 & 6.30 & G0III--IV & 1.13 & 0.49 & 118.96 & 6162 & 4.29  & +0.27  & 1.2  & GLTR \\
38629 & HD 38529 & 5.94 & G4V & 0.77 & 0.13 & 14.32 & 5646 & 3.92  & +0.37  & 1.2  & GLTR \\
145675 & 14 Her & 6.67 & K0V & 3.3 & 2.5 & 1650 & 5300 & 4.50  & +0.47  & 1.0  & FG \\ \hline
10700 & $\tau$ Cet & 3.50 & G8V & $\cdots$ & $\cdots$ & $\cdots$ & 5250 & 4.65  & $-$0.47$^{\ddagger}$  & 1.1  & AC \\
61421 & $\alpha$ CMi & 0.34 & F5IV--V & $\cdots$ & $\cdots$ & $\cdots$ & 6470 & 4.01  & $-$0.01  & 1.9  & F98 \\
186408 & 16 Cyg A & 5.96 & G1.5V & $\cdots$ & $\cdots$ & $\cdots$ & 5750 & 4.20  & +0.11  & 1.0  & G98 \\
102870 & $\beta$ Vir & 3.61 & F9V & $\cdots$ & $\cdots$ & $\cdots$ & 6085 & 4.04  & +0.14  & 1.4  & F98 \\
$\cdots$ & Sun & $\cdots$ & G2V & $\cdots$ & $\cdots$ & $\cdots$ & 5780 & 4.44  &~0.00  & 1.0  & SHKTT \\
\hline
\end{tabular}
\end{center}
Note. The $V$ magnitudes and the spectral types are from the SIMBAD
database. The data of the mass, the half-major axis, and the perid for
the planets around planet-harboring stars were taken from the 
``extrasolar planets'' home page 
(http://cfa-www.harvard.edu/planets/encycl.html).
The references for the $T_{\rm eff}$, $\log g$, [Fe/H] and $\xi$ values 
adopted in this paper are abbreviated in the last column: 
GL --- Gonzalez and Laws (2000), 
G98 --- Gonzalez (1998),
DS ---  Drake and Smith (1993),
GLTR --- Gonzalez et al. (2001),
GV --- Gonzalez and Vanture (1998),
FG --- Feltzing and Gonzalez (2001),
AC --- Arribas and Crivellari (1989),
F98 --- Fuhrmann (1998), and
SHKTT --- Sadakane et al. (1999).\\
$^{*}$ Chromospherically active star with spectroscopic anomalies
(cf. Drake, Smith 1993)\\
$^{\dagger}$ The companion mass may actually not be that of a planet,
because its orbital inclination angle ($i$) is suspected to be very small
(i.e., face-on system) according to Gatewood et al. (2001).\\
$^{\ddagger}$ The value from high-excitation Fe~{\sc i} lines was adopted 
(because of being less affected by a non-LTE effect).
\end{table}

\clearpage
\setcounter{table}{1}
\begin{table}[h]
\caption{Adopted $gf$ values of important lines.$^{*}$}
\begin{center}
\begin{tabular}{ccrc}\hline\hline
Species & $\lambda$ ($\rm\AA$) & $\chi$ (eV) & $\log gf$ \\
\hline
C~{\sc i} &5380.34 & 7.69 & $-$1.84   \\
S~{\sc i} &6052.58 & 7.87 & $-$1.33   \\
S~{\sc i} &6052.67 & 7.87 & $-$0.74   \\
V~{\sc i} &6081.44 & 1.05 & $-$0.58   \\
Co~{\sc i}&6082.42 & 3.51 & $-$0.52   \\
Fe~{\sc i}&6082.71 & 2.22 & $-$3.57   \\
Fe~{\sc ii}&6084.11& 3.20 & $-$3.81   \\
Ti~{\sc i}&6085.23 & 1.05 & $-$1.35   \\
Fe~{\sc i}&6085.26 & 2.76 & $-$3.21   \\
Ni~{\sc i}&6086.28 & 4.27 & $-$0.53   \\
Co~{\sc i}&6086.65 & 3.41 & $-$1.04   \\
Si~{\sc i}&6087.81 & 5.87 & $-$1.60   \\
O~{\sc i} &6156.74 &10.74 & $-$1.52   \\
O~{\sc i} &6156.76 &10.74 & $-$0.93   \\
O~{\sc i} &6156.78 &10.74 & $-$0.73   \\
Fe~{\sc i}&6157.73 & 4.08 & $-$1.26   \\
O~{\sc i} &6158.15 &10.74 & $-$1.89   \\
O~{\sc i} &6158.17 &10.74 & $-$1.03   \\
O~{\sc i} &6158.19 &10.73 & $-$0.44   \\
O~{\sc i} & 6300.30 & 0.00 & $-$9.82\\
Zn~{\sc i}&6362.34 & 5.80 & +0.15     \\
O~{\sc i}& 7771.94 & 9.15 &  +0.32 \\
O~{\sc i}& 7774.17 & 9.15 &  +0.17 \\
O~{\sc i}& 7775.39 & 9.15 & $-$0.05 \\
N~{\sc i}& 8683.40 &10.33 & $-$0.05 \\
S~{\sc i}& 8693.93 & 7.87 & $-$0.51 \\
S~{\sc i}& 8694.63 & 7.87 & +0.08   \\
C~{\sc i} &8727.12 & 1.26 & $-$8.21   \\
Fe~{\sc i}&8727.13 & 4.19 & $-$3.93   \\
Si~{\sc i}&8728.01 & 6.18 & $-$0.61   \\
Si~{\sc i}&8728.59 & 6.18 & $-$1.72   \\
Fe~{\sc i}&8729.15 & 3.42 & $-$2.95   \\
Si~{\sc i}&8729.28 & 6.18 & $-$2.30   \\
\hline
\end{tabular}
\end{center}
$^{*}$ All data were taken from the compilation of Kurucz and Bell (1995).
\end{table}

\clearpage
\setcounter{table}{2}
\begin{table}[h]
\caption{Data of measured equivalent widths (in m$\rm\AA$).}
\begin{center}
\scriptsize
\begin{tabular}{crrrrrrrrrr}\hline\hline
Star & C5380 & S6052 & O6300 & Zn6362 & O7771 & O7774 & O7775 & N8683 & S8693 & S8694 \\ \hline
$\tau$ Boo & 56.8  & 35.9  & $\cdots$  & 29.6  & 158.8  & 138.2  & 125.2  & 36.2  & 52.0  & 85.1 \\
$\upsilon$ And & 41.5  & 25.1  & 4.6  & 24.5  & 122.9  & 115.8  & 96.4  & 19.6  & 29.1  & 55.1 \\
16 Cyg B & 23.9  & 11.2  & 6.5  & 18.8  & 70.3  & 60.7  & 54.4  & 10.8  & 11.8  & 30.8 \\
$\epsilon$ Eri & 9.8  & 5.5  & 5.2  & 9.3  & $\cdots$  & $\cdots$  & $\cdots$  & $\cdots$  & $\cdots$  & $\cdots$ \\
47 UMa & 25.8  & 13.8  & 6.0  & 22.8  & 82.8  & 79.5  & 60.6  & 13.0  & 13.8  & 33.2 \\
HD 89744 & 43.0  & 28.5  & 4.9  & 26.9  & 137.3  & 131.0  & 111.3  & 22.9  & 29.9  & 63.7 \\
51 Peg & 29.5  & 18.2  & 7.2  & 25.4  & $\cdots$  & $\cdots$  & $\cdots$  & $\cdots$  & $\cdots$  & $\cdots$ \\
55 Cnc & 19.5  & 20.1  & 9.9  & 26.6  & 45.2  & 49.9  & 31.1  & 10.5  & 9.0  & 20.3 \\
70 Vir & 14.9  & 10.4  & 9.5  & 19.3  & 62.9  & 54.9  & 42.2  & 4.0  & 13.3  & 19.5 \\
$\rho$ CrB & 15.9  & 10.8  & 5.4  & 19.1  & 72.8  & 73.0  & 56.4  & 6.1  & 11.4  & 27.8 \\
HD 217107 & 29.4  & 23.6  & 8.7  & 26.8  & $\cdots$  & $\cdots$  & $\cdots$  & $\cdots$  & $\cdots$  & $\cdots$ \\
HD 52265 & 37.1  & 25.5  & 7.4  & 26.1  & 118.8  & 111.3  & 88.0  & 16.6  & 27.4  & 54.3 \\
HD 38529 & 35.5  & 27.0  & 12.5  & 36.2  & 85.5  & 82.8  & 63.7  & 17.0  & 30.5  & 42.0 \\
14 Her & 23.9  & 23.7  & 8.4  & 29.2  & $\cdots$  & $\cdots$  & $\cdots$  & $\cdots$  & $\cdots$  & $\cdots$ \\
\hline
$\tau$ Cet & 6.7  & 4.8  & 5.0  & 10.7  & $\cdots$  & $\cdots$  & $\cdots$  & $\cdots$  & $\cdots$  & $\cdots$ \\
$\alpha$ CMi & 51.6  & 24.9  & 3.6  & 15.9  & 170.4  & 164.1  & 136.5  & 30.2  & 28.7  & 59.3 \\
16 Cyg A & 24.6  & 15.3  & 7.4  & 24.4  & 79.6  & 66.5  & 59.4  & 8.0  & 16.0  & 32.5 \\
$\beta$ Vir & 38.7  & 22.8  & 6.6  & 26.3  & $\cdots$  & $\cdots$  & $\cdots$  & $\cdots$  & $\cdots$  & $\cdots$ \\
Sun & 19.9  & 11.2  & 5.2  & 19.7  & 72.4  & 62.7  & 47.9  & 6.4  & 10.6  & 27.3 \\
\hline
\end{tabular}
\end{center}
\end{table}

\clearpage
\setcounter{table}{3}
\begin{table}[h]
\caption{Comparison of the abundances from different lines.}
\scriptsize
\begin{center}
\begin{tabular}
{c
c@{ }c
c@{ }c@{ }c@{ }c
c@{ }c
c@{ }c
c@{ }c@{ }c}\hline\hline
Star 
& [C]$_{5380}$ & [C]$_{8727}^{\rm fit}$ 
& [O]$_{6158}^{\rm fit}$ & [O]$_{6300}$ & [O]$_{7773}$ & $\Delta_{7773}$ 
& [Si]$_{6087}^{\rm fit}$ & [Si]$_{8728}^{\rm fit}$ 
& [S]$_{6052}$ & [S]$_{8694}$ 
& [Fe]$_{6085}^{\rm fit}$ & [Fe]$_{6157}^{\rm fit}$ 
& [Fe]$_{8729}^{\rm fit}$ \\
\hline
$\tau$ Boo & +0.31 & +0.42 & +0.09 & $\cdots$ & +0.19 & ($-$0.38)  & +0.26 & +0.38 & +0.28 & +0.59 & +0.18 & +0.21 & +0.27 \\
$\upsilon$ And & +0.19 & +0.29 & +0.07 & $-$0.08 & +0.14 & ($-$0.27)  & +0.18 & +0.24 & +0.16 & +0.28 & +0.10 & +0.09 & +0.11 \\
16 Cyg B & +0.13 & $-$0.02 & +0.06 & +0.07 & +0.07 & ($-$0.11)  & +0.09 & +0.10 & +0.02 & +0.08 & +0.05 & +0.05 & +0.09 \\
$\epsilon$ Eri & +0.14 & $\cdots$ & $\cdots$ & +0.00 & $\cdots$ & $\cdots$  & $-$0.04 & $\cdots$ & +0.17 & $\cdots$ & $-$0.10 & $\cdots$ & $\cdots$ \\
47 UMa & +0.10 & +0.05 & +0.05 & $-$0.01 & +0.11 & ($-$0.15)  & +0.01 & +0.05 & +0.04 & +0.08 & $-$0.08 & $-$0.05 & $-$0.01 \\
HD 89744 & +0.13 & +0.34 & +0.02 & +0.07 & +0.10 & ($-$0.31)  & +0.26 & +0.27 & +0.16 & +0.21 & +0.21 & +0.15 & +0.35 \\
51 Peg & +0.26 & $\cdots$ & +0.20 & +0.20 & $\cdots$ & $\cdots$  & +0.16 & $\cdots$ & +0.26 & $\cdots$ & +0.19 & +0.21 & $\cdots$ \\
55 Cnc & +0.33 & $-$0.34 & $\cdots$ & +0.37 & +0.29 & ($-$0.06)  & +0.26 & +0.50 & +0.69 & +0.33 & +0.27 & $\cdots$ & +0.43 \\
70 Vir & $-$0.17 & $-$0.11 & $-$0.07 & $-$0.01 & $-$0.02 & ($-$0.13)  & $-$0.01 & $-$0.02 & $-$0.04 & +0.15 & $-$0.08 & $-$0.06 & $-$0.08 \\
$\rho$ CrB & $-$0.19 & $-$0.25 & +0.08 & $-$0.23 & $-$0.02 & ($-$0.15)  & $-$0.13 & $-$0.17 & $-$0.09 & $-$0.05 & $-$0.30 & $-$0.29 & $-$0.24 \\
HD 217107 & +0.36 & $\cdots$ & +0.27 & +0.32 & $\cdots$ & $\cdots$  & +0.27 & $\cdots$ & +0.52 & $\cdots$ & +0.28 & +0.26 & $\cdots$ \\
HD 52265 & +0.15 & +0.33 & +0.09 & +0.26 & +0.14 & ($-$0.24)  & +0.24 & +0.43 & +0.21 & +0.28 & +0.19 & +0.26 & +0.27 \\
HD 38529 & +0.31 & +0.25 & +0.23 & +0.30 & +0.19 & ($-$0.19)  & +0.35 & +0.37 & +0.42 & +0.60 & +0.37 & +0.41 & +0.39 \\
14 Her & +0.46 & $\cdots$ & $\cdots$ & +0.35 & $\cdots$ & $\cdots$  & +0.36 & $\cdots$ & +0.79 & $\cdots$ & +0.39 & $\cdots$ & $\cdots$ \\
\hline
$\tau$ Cet & $-$0.13 & $\cdots$ & $\cdots$ & $-$0.18 & $\cdots$ & $\cdots$  & $-$0.21 & $\cdots$ & +0.03 & $\cdots$ & $-$0.58 & $\cdots$ & $\cdots$ \\
$\alpha$ CMi & +0.15 & +0.26 & +0.07 & $-$0.16 & +0.17 & ($-$0.43)  & $-$0.05 & +0.04 & +0.00 & +0.07 & $-$0.08 & $-$0.17 & +0.07 \\
16 Cyg A & +0.06 & $-$0.38 & +0.04 & +0.09 & +0.06 & ($-$0.14)  & +0.09 & +0.12 & +0.10 & +0.18 & +0.07 & +0.12 & +0.10 \\
$\beta$ Vir & +0.14 & $\cdots$ & +0.10 & +0.04 & $\cdots$ & $\cdots$  & +0.16 & $\cdots$ & +0.11 & $\cdots$ & +0.11 & $\cdots$ & $\cdots$ \\
\hline
Sun & 8.67 & 8.41 & 8.87 & 8.97 & 8.83 & ($-$0.11)  & 7.46 & 7.84 & 7.36 & 7.19  & 7.54 & 7.62 & 7.64 \\
\hline
\end{tabular}
\end{center}
Note. The presented stellar abundances are the differential ones
([X] $\equiv \log\epsilon_{\rm X} - \log\epsilon_{\odot}$)
relative to the solar abundance given in the last row 
($\log\epsilon_{\odot}$; expressed with the usual normalization of 
$\log\epsilon_{\rm H}=12.00$)
. 
The values in the 7th column are the non-LTE corrections
($\Delta \equiv \log\epsilon_{\rm NLTE} - \log\epsilon_{\rm LTE}$)
for the O~{\sc i} 7771--5 triplet (mean of the three lines)
showing an appreciable non-LTE effect, while LTE was assumed for the
other two oxygen lines (O~{\sc i} 6158 and [O~{\sc i}] 6300)
for which this correction is negligibly small (cf. Takeda 2001).
\end{table}

\clearpage
\setcounter{table}{4}
\begin{table}[h]
\caption{Abundance variations to changes in the model parameters.}
\scriptsize
\begin{center}
\begin{tabular}
{c
c@{ }c@{ }c
c@{ }c@{ }c
}\hline\hline
  & $\Delta_{T}^{*}$ & $\Delta_{g}^{\dag}$ & $\Delta_{\xi}^{\ddag}$ & $\Delta_{T}^{*}$ & $\Delta_{g}^{\dag}$ & $\Delta_{\xi}^{\ddag}$ \\
  & \multicolumn{3}{c}{$\upsilon$ And} & \multicolumn{3}{c}{70 Vir} \\
  & \multicolumn{3}{c}{(6140 K, 4.12, 1.4 km~s$^{-1}$)} & \multicolumn{3}{c}{(5500 K, 3.90, 1.0 km~s$^{-1}$)} \\
\hline
\multicolumn{7}{c}{(Analysis of individual lines)} \\
C~{\sc i} 5380 & $-$0.07  & +0.09  & $-$0.02  & $-$0.09  & +0.10  & +0.00 \\
N~{\sc i} 8683 & $-$0.11  & +0.10  & +0.01  & $-$0.13  & +0.10  & +0.02 \\
O~{\sc i} 6300 & +0.04  & +0.12  & +0.00  & +0.03  & +0.13  & +0.00 \\
O~{\sc i} 7771 & $-$0.11  & +0.05  & $-$0.08  & $-$0.16  & +0.08  & $-$0.05 \\
O~{\sc i} 7774 & $-$0.12  & +0.05  & $-$0.08  & $-$0.16  & +0.09  & $-$0.05 \\
O~{\sc i} 7775 & $-$0.11  & +0.07  & $-$0.07  & $-$0.15  & +0.10  & $-$0.04 \\
S~{\sc i} 6052 & $-$0.06  & +0.08  & $-$0.01  & $-$0.09  & +0.10  & +0.00 \\
S~{\sc i} 8693 & $-$0.08  & +0.08  & $-$0.02  & $-$0.11  & +0.10  & $-$0.01 \\
S~{\sc i} 8694 & $-$0.08  & +0.07  & $-$0.04  & $-$0.11  & +0.09  & $-$0.02 \\
Zn~{\sc i} 6362 & +0.03  & +0.04  & $-$0.03  & $-$0.03  & +0.07  & $-$0.03 \\
\hline
\multicolumn{7}{c}{(6080--6089 $\rm\AA$ fitting analysis)} \\
Si & +0.06  & +0.01  & +0.00  & +0.03  & +0.05  & +0.00 \\
Ti & +0.23  & $-$0.11  & +0.00  & +0.22  & $-$0.04  & $-$0.05 \\
V & +0.11  & +0.00  & $-$0.01  & +0.16  & $-$0.01  & $-$0.03 \\
Fe & +0.05  & +0.06  & $-$0.04  & +0.08  & +0.04  & $-$0.07 \\
Co & +0.15  & $-$0.06  & +0.00  & +0.10  & +0.02  & $-$0.02 \\
Ni & +0.08  & +0.00  & $-$0.06  & +0.07  & +0.03  & $-$0.09 \\
\hline
\multicolumn{7}{c}{(6156.5--6158.5 $\rm\AA$ fitting analysis)} \\
O & $-$0.12  & +0.11  & +0.00  & $-$0.14  & +0.10  & $-$0.01 \\
Fe & +0.09  & $-$0.01  & $-$0.11  & +0.10  & +0.00  & $-$0.19 \\
\hline
\multicolumn{7}{c}{(8726--8730.5 $\rm\AA$ fitting analysis)} \\
C & +0.01  & +0.11  & +0.00  & $-$0.05  & +0.15  & $-$0.01 \\
Si & +0.04  & $-$0.03  & $-$0.07  & $-$0.01  & +0.01  & $-$0.07 \\
Fe & +0.09  & +0.01  & +0.02  & +0.11  & +0.00  & $-$0.03 \\
\hline
\end{tabular}
\end{center}
$^{*}$ $\Delta T_{\rm eff}$ = +150 K.\\
$^{\dag}$ $\Delta\log g$ = +0.3.\\
$^{\ddag}$ $\Delta\xi$ = +0.5 km~s$^{-1}$.\\
\end{table}

\clearpage
\setcounter{table}{5}
\begin{table}[h]
\caption{Final results of the abundances.}
\scriptsize
\begin{center}
\begin{tabular}
{c
c@{ }c@{ }c@{ }c@{ }c
c@{ }c@{ }c@{ }c@{ }c@{ }c
r@{ }r}\hline\hline
  & [C] & [N] & [O] & [S] & [Zn] & [Si] & [Fe] & [Co] & [Ni] & [V] & [Ti] & \multicolumn{2}{c}{Slope $a^{*}$} \\
$T_{\rm c}$ (K) & 75  & 120  & 180  & 648  & 660  & 1311  & 1336  & 1351  & 1354  & 1450  & 1549  & with & without \\
  & \multicolumn{5}{c}{Volatile elements} & \multicolumn{6}{c}{Refractory elements} & N & N \\
\hline
$\tau$ Boo & +0.35 & +0.47 & +0.14 & +0.44 & +0.29 & +0.26 & +0.18 & +0.42 & +0.24 & +0.10 & +0.22 & $-$8.45  & $-$4.58 \\
$\upsilon$ And & +0.22 & +0.24 & +0.07 & +0.22 & +0.07 & +0.18 & +0.10 & +0.26 & +0.12 & +0.14 & +0.08 & $-$2.78  & $-$0.76 \\
16 Cyg B & +0.08 & +0.31 & +0.07 & +0.05 & $-$0.02 & +0.09 & +0.05 & +0.12 & +0.10 & +0.07 & +0.00 & $-$5.19  & +0.74 \\
$\epsilon$ Eri & +0.14 & $\cdots$ & +0.00 & +0.17 & $-$0.15 & $-$0.04 & $-$0.10 & +0.00 & $-$0.13 & +0.08 & $-$0.05 & $\cdots$ & $-$8.16 \\
47 UMa & +0.08 & +0.31 & +0.06 & +0.06 & +0.06 & +0.01 & $-$0.08 & $-$0.01 & $-$0.04 & $-$0.05 & $-$0.11 & $-$15.81  & $-$10.77 \\
HD 89744 & +0.20 & +0.19 & +0.06 & +0.19 & +0.17 & +0.26 & +0.21 & +0.38 & +0.16 & +0.27 & +0.24 & +7.99  & +9.32 \\
51 Peg & +0.26 & $\cdots$ & +0.20 & +0.26 & +0.20 & +0.16 & +0.19 & +0.30 & +0.29 & +0.12 & +0.12 & $\cdots$ & $-$3.83 \\
55 Cnc & +0.11 & +0.71 & +0.32 & +0.51 & +0.51 & +0.26 & +0.27 & +0.55 & +0.50 & +0.46 & +0.18 & $-$3.06  & +6.15 \\
70 Vir & $-$0.15 & $-$0.17 & $-$0.04 & +0.05 & $-$0.08 & $-$0.01 & $-$0.08 & +0.02 & $-$0.09 & $-$0.05 & $-$0.10 & +3.93  & +1.50 \\
$\rho$ CrB & $-$0.21 & $-$0.05 & $-$0.02 & $-$0.07 & $-$0.13 & $-$0.13 & $-$0.30 & $-$0.15 & $-$0.30 & $-$0.21 & $-$0.18 & $-$9.42  & $-$8.32 \\
HD 217107 & +0.36 & $\cdots$ & +0.29 & +0.52 & +0.31 & +0.27 & +0.28 & +0.49 & +0.41 & +0.33 & +0.20 & $\cdots$ & $-$2.69 \\
HD 52265 & +0.21 & +0.15 & +0.14 & +0.25 & +0.17 & +0.24 & +0.19 & +0.39 & +0.25 & +0.26 & +0.28 & +7.85  & +7.28 \\
HD 38529 & +0.29 & +0.46 & +0.23 & +0.51 & +0.39 & +0.35 & +0.37 & +0.58 & +0.45 & +0.48 & +0.31 & +5.79  & +9.12 \\
14 Her & +0.46 & $\cdots$ & +0.35 & +0.79 & +0.56 & +0.36 & +0.39 & +0.67 & +0.59 & +0.58 & +0.35 & $\cdots$ & +0.11 \\
\hline
$\tau$ Cet & $-$0.13 & $\cdots$ & $-$0.18 & +0.03 & $-$0.21 & $-$0.21 & $-$0.58 & $-$0.33 & $-$0.51 & $-$0.43 & $-$0.45 & $\cdots$ & $-$25.44 \\
$\alpha$ CMi & +0.19 & +0.27 & +0.06 & +0.03 & $-$0.12 & $-$0.05 & $-$0.08 & +0.10 & $-$0.13 & $-$0.07 & $-$0.25 & $-$19.31  & $-$15.45 \\
16 Cyg A & $-$0.09 & +0.04 & +0.06 & +0.14 & +0.11 & +0.09 & +0.07 & +0.16 & +0.14 & +0.12 & +0.09 & +7.40  & +7.85 \\
$\beta$ Vir & +0.14 & $\cdots$ & +0.08 & +0.11 & +0.10 & +0.16 & +0.11 & +0.22 & +0.13 & +0.09 & +0.10 & $\cdots$ & +1.77 \\
\hline
Sun & 8.58 & 8.13  & 8.87 & 7.28 & 4.51 & 7.46 & 7.54 & 4.72 & 6.35 & 3.89 & 5.01 & ------ & ------ \\
\hline
\end{tabular}
\end{center}
Note. The abundance results presented here are the differential values
relative to the solar abundances (given in the last row).\\
$^{*}$ The slope ($a$; expressed in unit of $10^{-5}$~dex~K$^{-1}$) 
of the linear-regression line, [X/H]=$a \cdot T_{\rm c} + b$.
\end{table}

%\clearpage
\onecolumn

\begin{figure}
  \begin{center}
    \FigureFile(155mm,155mm){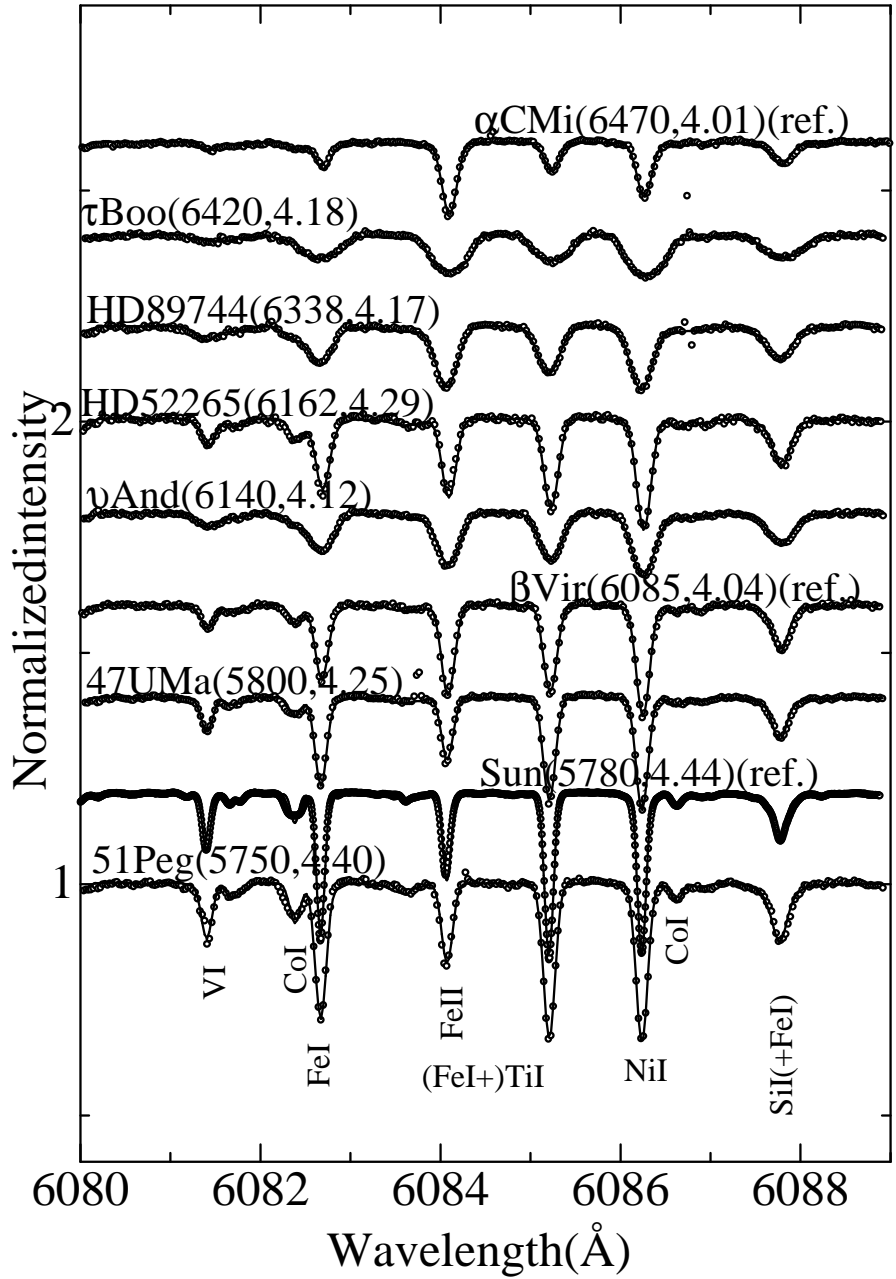}
    %%% \FigureFile(width,height){filename}
  \end{center}
%  \caption{This is the first figure.}\label{fig:sample}
%Figure 1a
\caption{
Spectral fitting at the 6080--6089 $\rm\AA$ region
to determine the abundances of Si, Ti, V, Fe, Co, and Ni.
Open circles denote the observed data, while the solid curves
represent the best-fit theoretical spectra. Each of the spectra,
arranged in the decreasing order of $T_{\rm eff}$, are vertically offset
by 0.2 with respect to the adjacent one.\\
(a) Higher $T_{\rm eff}$ stars with 5750~K $\ltsim T_{\rm eff} \ltsim$ 6500~K.
}
\end{figure}

\setcounter{figure}{0}
\begin{figure}
  \begin{center}
    \FigureFile(155mm,155mm){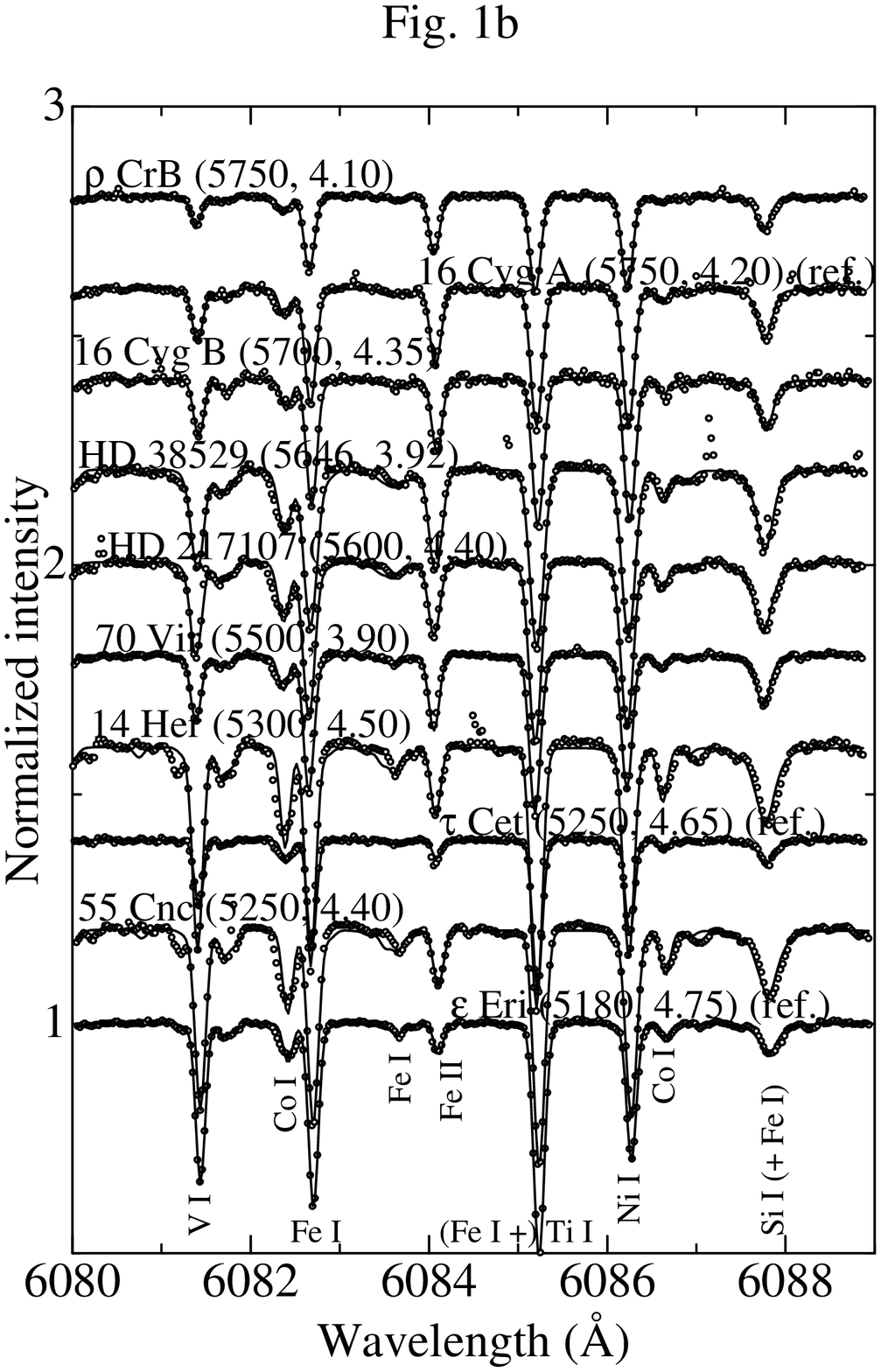}
    %%% \FigureFile(width,height){filename}
  \end{center}
%  \caption{This is the first figure.}\label{fig:sample}
%Figure 1b
\caption{
(b) Lower $T_{\rm eff}$ stars with 5200~K $\ltsim T_{\rm eff} \ltsim$ 5750~K.
}
\end{figure}

\begin{figure}
  \begin{center}
    \FigureFile(160mm,180mm){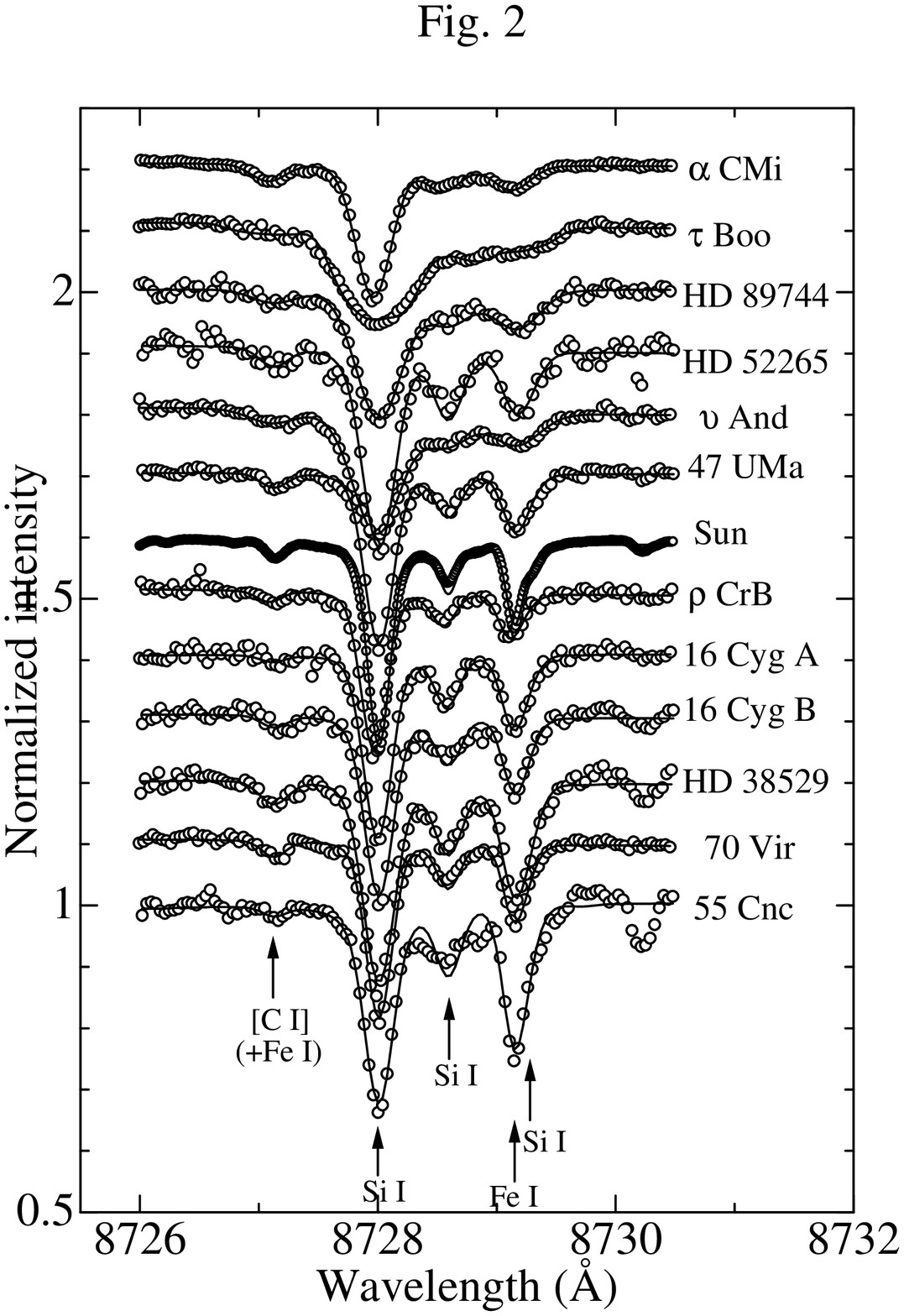}
    %%% \FigureFile(width,height){filename}
  \end{center}
%  \caption{This is the first figure.}\label{fig:sample}
%Figure 2
\caption{Spectra fitting at the 8726--8730.5 $\rm\AA$ region
to determine the abundances of C, Si, and Fe. A vertical offset of 0.1
is applied to each spectra relative to the adjacent one.
Otherwise, the same as in figure 1.}
\end{figure}

\begin{figure}
  \begin{center}
    \FigureFile(160mm,160mm){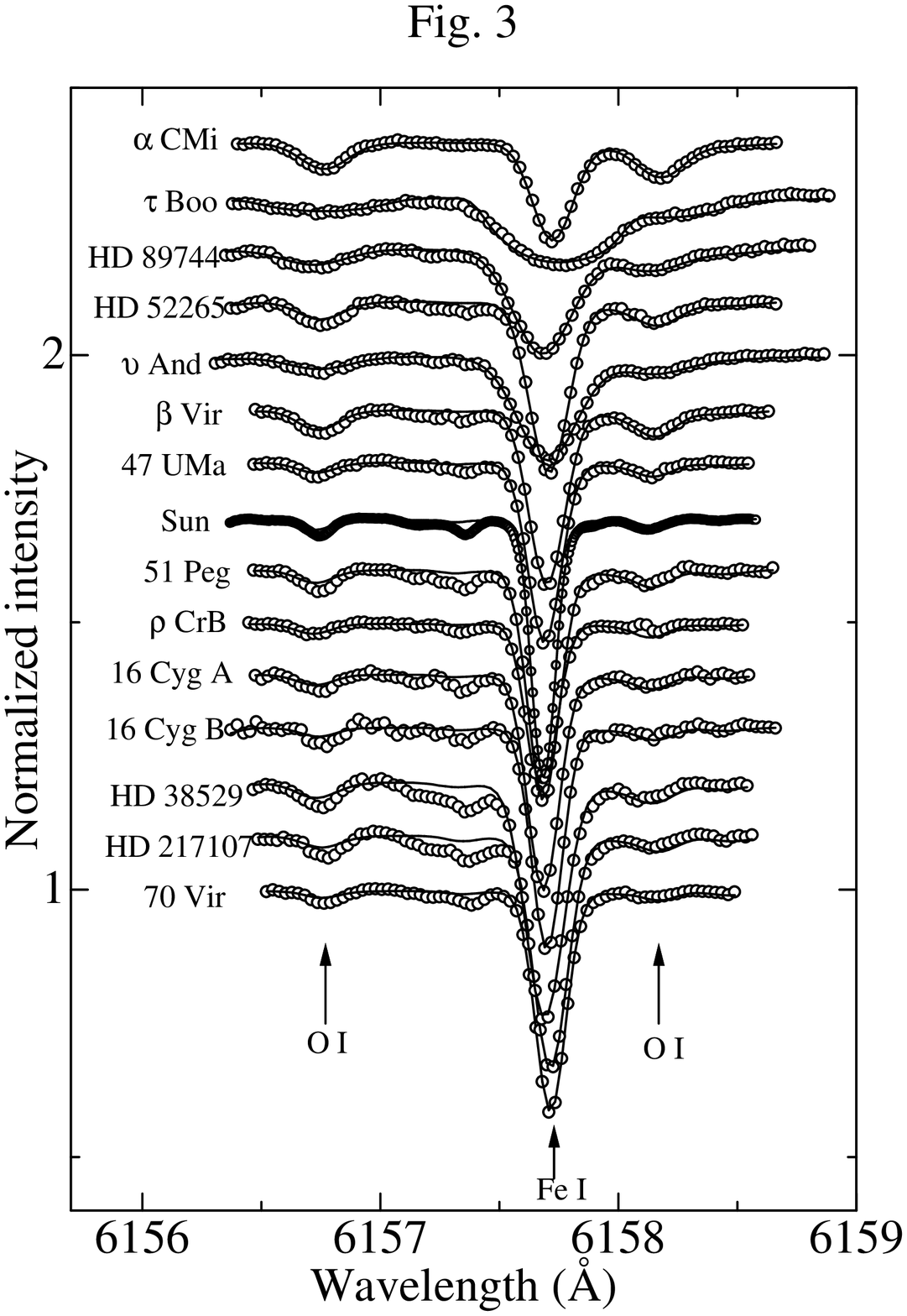}
    %%% \FigureFile(width,height){filename}
  \end{center}
%  \caption{This is the first figure.}\label{fig:sample}
%Figure 3
\caption{Spectra fitting at the 6156.5--6158.5 $\rm\AA$ region
to determine the abundances of O and Fe. A vertical offset of 0.1
is applied to each spectra relative to the adjacent one.
Otherwise, the same as in figure 1.}
\end{figure}

\begin{figure}
  \begin{center}
    \FigureFile(160mm,160mm){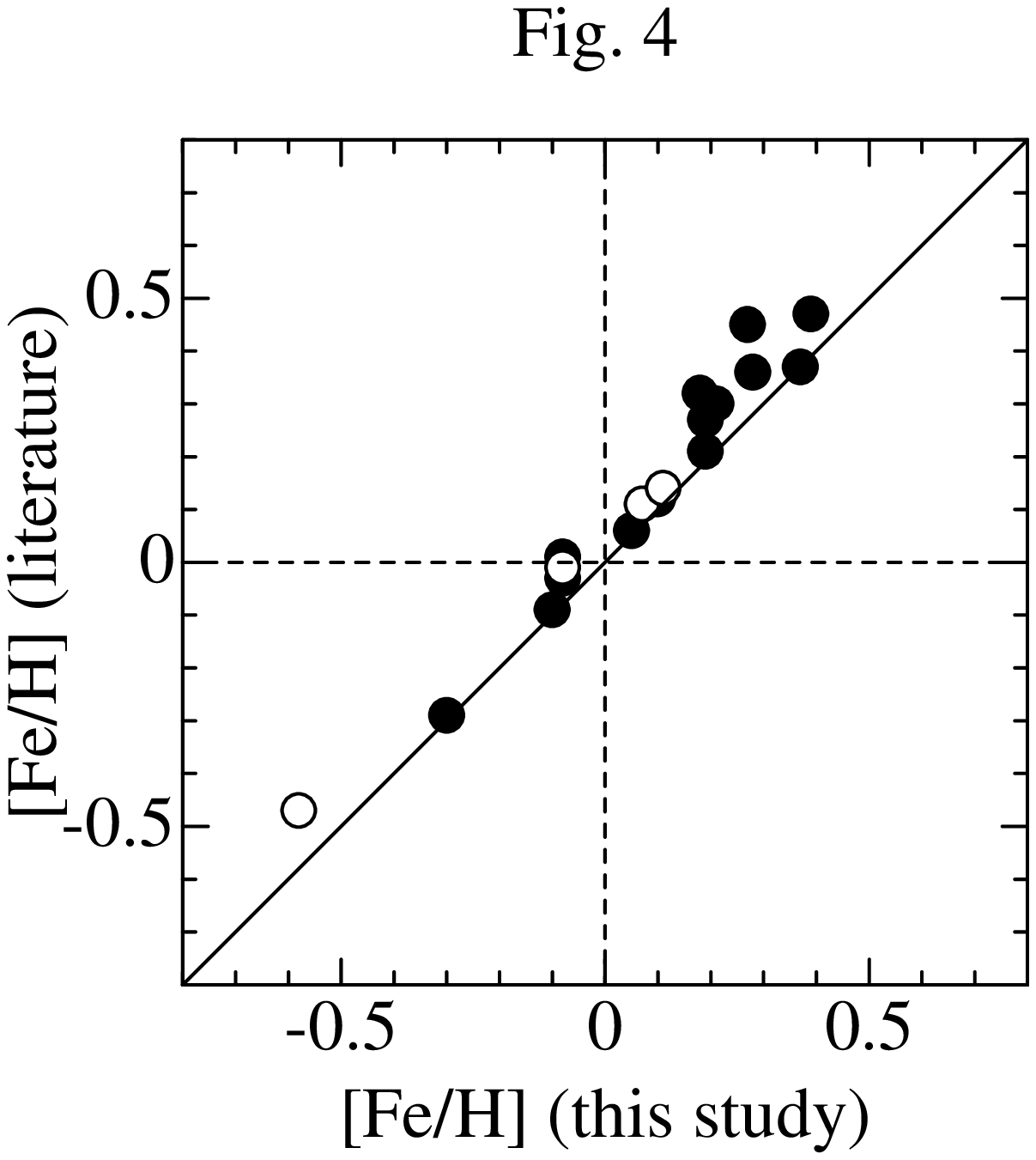}
    %%% \FigureFile(width,height){filename}
  \end{center}
%  \caption{This is the first figure.}\label{fig:sample}
%Figure 4
\caption{Comparison of the [Fe/H] values derived in this study
with those taken from various literatures (cf. table 1).
Filled circles correspond to planet-harboring stars, while
open circles to the reference stars.}
\end{figure}

\begin{figure}
  \begin{center}
    \FigureFile(160mm,160mm){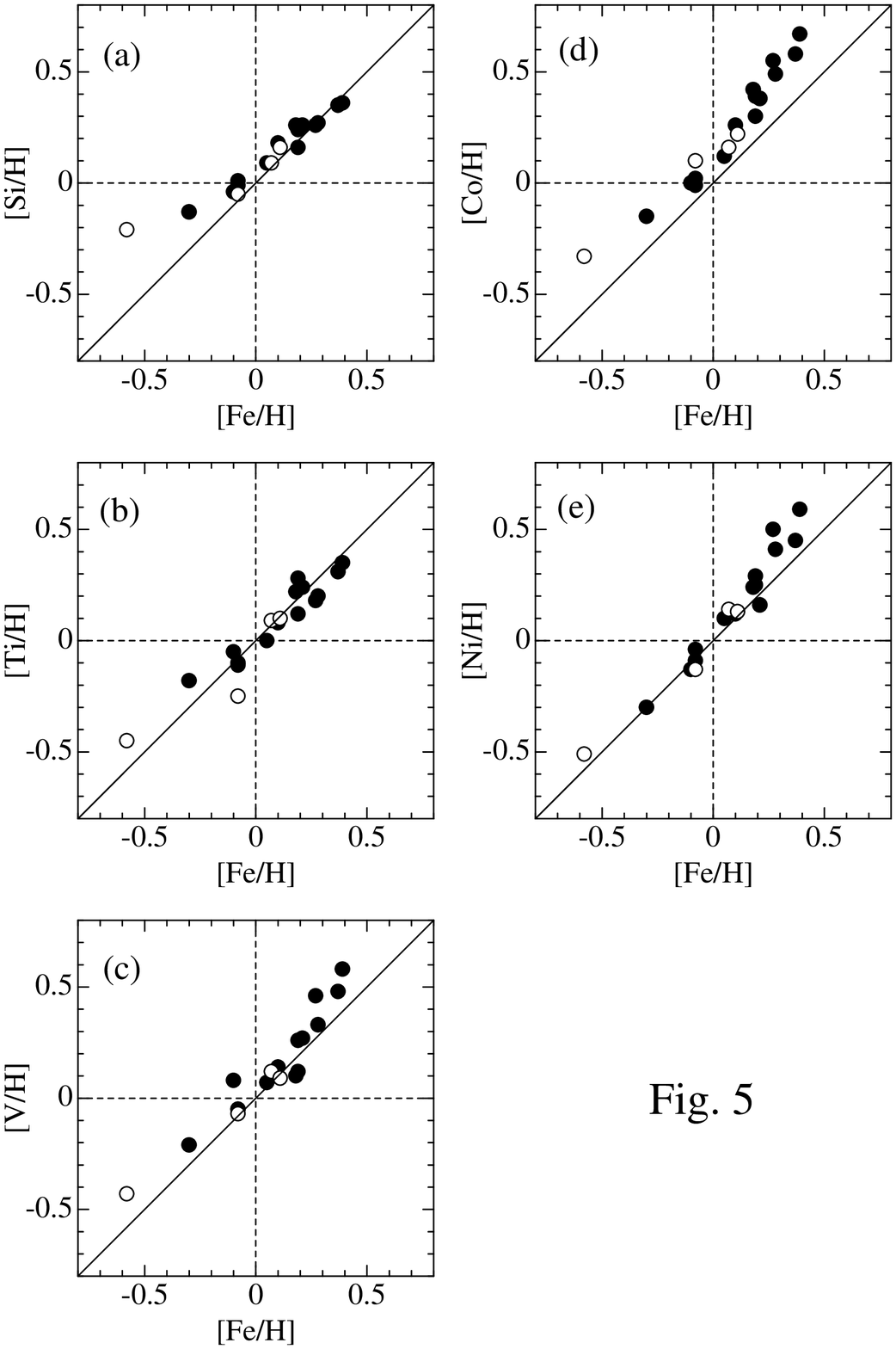}
    %%% \FigureFile(width,height){filename}
  \end{center}
%  \caption{This is the first figure.}\label{fig:sample}
%Figure 5
\caption{Comparison of the abundances of five refractory elements
with the Fe abundance: (a) Si, (b) Ti, (c) V, (d) Co, and (e) Ni.
The same meanings of the symbols as in figure 4.}
\end{figure}

\begin{figure}
  \begin{center}
    \FigureFile(160mm,160mm){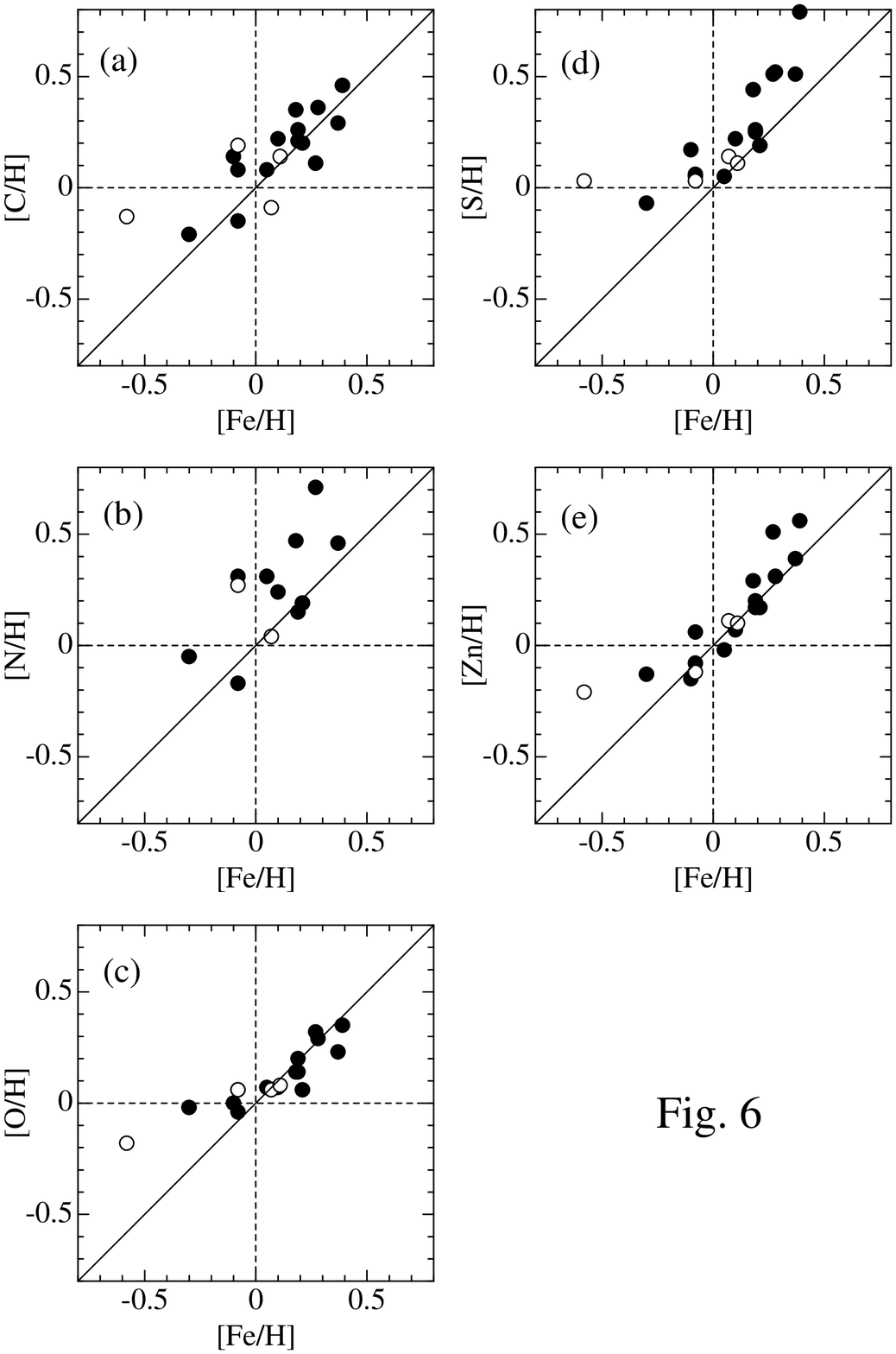}
    %%% \FigureFile(width,height){filename}
  \end{center}
%  \caption{This is the first figure.}\label{fig:sample}
%Figure 6
\caption{Comparison of the abundances of five volatile elements
with the Fe abundance: (a) C, (b) N, (c) O, (d) S, and (e) Zn.
The same meanings of the symbols as in figure 4.}
\end{figure}

\begin{figure}
  \begin{center}
    \FigureFile(160mm,160mm){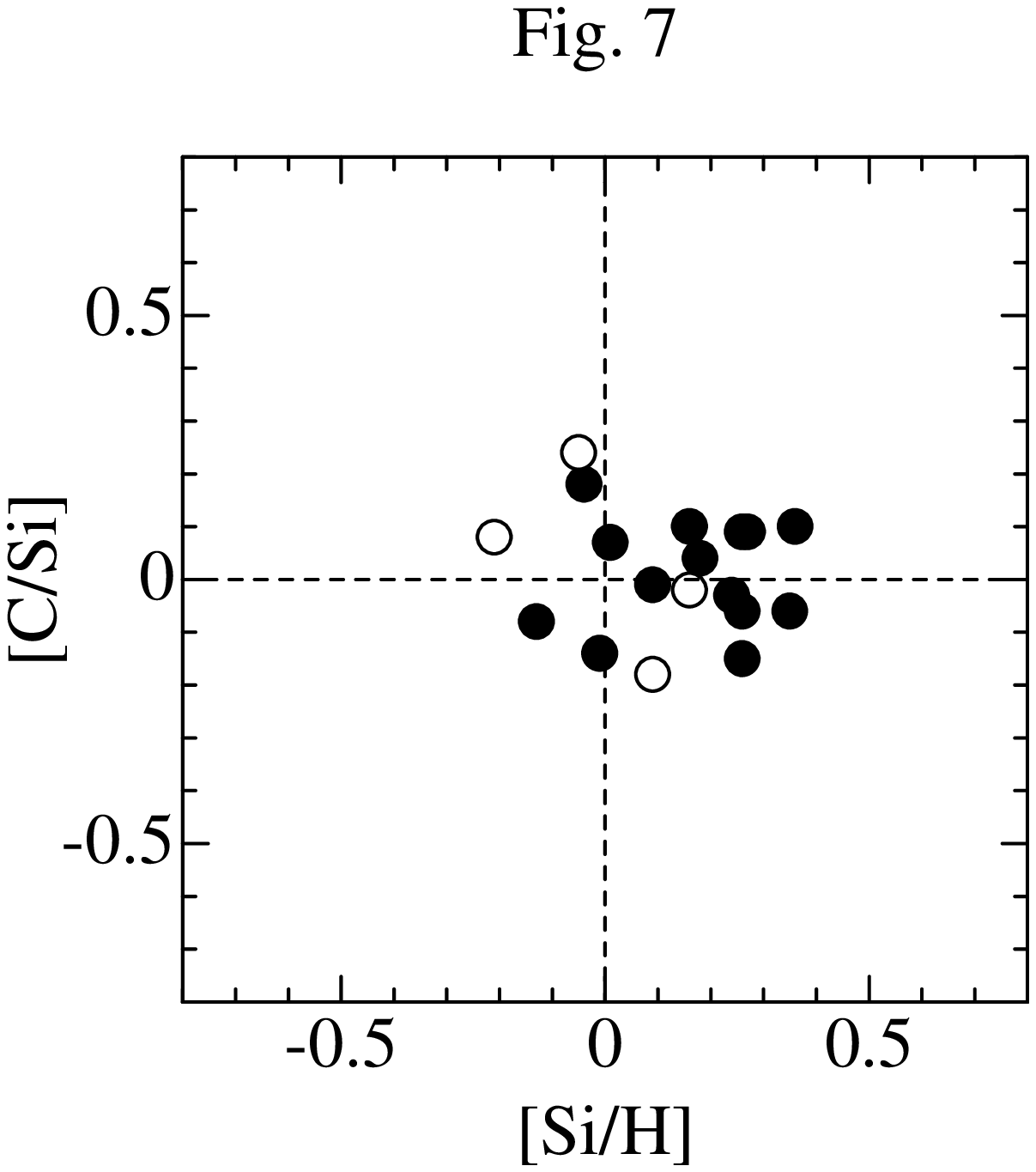}
    %%% \FigureFile(width,height){filename}
  \end{center}
%  \caption{This is the first figure.}\label{fig:sample}
%Figure 7
\caption{Comparison of [C/Si] with [Si/H].
The same meanings of the symbols as in figure 4.}
\end{figure}
\end{document}